\documentclass[preprint]{article}
\usepackage{authblk}
\usepackage{graphicx}
\usepackage{amsmath,amssymb,amsthm}
\usepackage{mathrsfs}
\usepackage{slashed}
\usepackage{dcolumn}
\usepackage{multirow}
\usepackage{bm}
\usepackage{appendix}
\usepackage{makecell}
\allowdisplaybreaks[4]
\usepackage[colorlinks,citecolor=blue,urlcolor=blue,linkcolor=blue]{hyperref}
\usepackage[numbers,sort&compress]{natbib}
\usepackage[left=0.8in,right=0.8in,top=1in,bottom=1in]{geometry}
\begin{document}
	\title{Light-cone Sum Rule Analysis of Semileptonic Decays $\Lambda_b^0 \to \Lambda_c^+ \ell^- \overline{\nu}_\ell$}
	\author{Hui-Hui Duan\footnote{duanhuihui19@nudt.edu.cn}}
	\author{Yong-Lu Liu\footnote{yongluliu@nudt.edu.cn}}
	\author{Ming-Qiu Huang\footnote{Corresponding author: mqhuang@nudt.edu.cn}}
	\affil{Department of Physics, National University of Defense Technology, Changsha 410073, Hunan, People's Republic of China}
	\renewcommand*{\Affilfont}{\small\it}
    \date{}
	\maketitle

	\begin{abstract}  
		  	
    	In this work, we analyze the semileptonic decay processes of $\Lambda_b \to \Lambda_c$ in the light-cone sum rule approach. In order to calculate the form factors of the $\Lambda_b$ baryon transition matrix element,  we use the light-cone distribution amplitudes of $\Lambda_b$ obtained from the QCD sum rule in the heavy quark effective field theory framework. With the calculation of the six form factors of the $\Lambda_b \to \Lambda_c$ transition matrix element, the differential decay widths of $\Lambda_b^0 \to \Lambda_c^+ \ell^- \overline{\nu}_\ell (\ell = e, ~\mu, ~\tau)$ and their absolute branching fractions are obtained. Additionally, the ratio of $R(\Lambda_c^+) \equiv \mathcal{B}r(\Lambda_b^0 \to \Lambda_c^+ \tau^- \overline{\nu}_\tau)/\mathcal{B}r(\Lambda_b^0 \to \Lambda_c^+ \mu^- \overline{\nu}_\mu)$ is also obtained in this work. Our results are in accord with the newest experimental result and other theoretical calculations and predictions.
    \end{abstract}
    
    \section{Introduction}\label{sec:I}
   
    Semileptonic decays of $\Lambda_b$ to $\Lambda_c$ baryons provide a good way to study and determine the CKM matrix element $|V_{cb}|$. In most cases, the CKM matrix element $|V_{cb}|$ is extracted from the $B$ to $D$ meson semileptonic decay processes. Compared with different methods and models, there is a small difference in the value of $|V_{cb}|$. Meanwhile, theoretical analysis and experimental measurements of the ratio $\mathcal{R}(\Lambda_c^+) \equiv \mathcal{B}r(\Lambda_b^0 \to \Lambda_c^+ \tau^- \overline{\nu}_\tau)/\mathcal{B}r(\Lambda_b^0 \to \Lambda_c^+ \mu^- \overline{\nu}_\mu)$ also give a good test of the Standard Model and lepton universality.
     
     Experimentally, LHCb reported their measurement of the branching fractions of semileptonic decay $\Lambda_b$ to $\Lambda_c$ with tau lepton final state with a significance of $6.1\sigma$ recently \cite{LHCb:2022piu}. They gave the branching fraction of the semi-tauonic decay $\Lambda_b^0\to\Lambda_c^+\tau^-\overline{\nu}_\tau$,  $\mathcal{B}r(\Lambda_b^0\to\Lambda_c^+\tau^-\overline{\nu}_\tau)=(1.50\pm0.16_{stat}\pm 0.25_{syst}\pm 0.23)\%$. At the same time, the experimental test also gave the ratio of branching fractions of the processes $\Lambda_b^0\to\Lambda_c^+\tau^-\overline{\nu}_\tau$ and  $\Lambda_b^0\to\Lambda_c^+\mu^-\overline{\nu}_\mu$, which is $\mathcal{R}(\Lambda_c^+) \equiv \mathcal{B}r(\Lambda_b^0\to\Lambda_c^+\tau^-\overline{\nu}_\tau)/\mathcal{B}r(\Lambda_b^0\to\Lambda_c^+\mu^-\overline{\nu}_\mu)=(0.242\pm0.026\pm0.040\pm0.059)$, with the last uncertainty coming from the errors of measurement of $\Lambda_b^0\to\Lambda_c^+\mu^-\overline{\nu}_\mu$ branching fraction. In the early experiments, the DELPHI and CDF collaborations reported the branching fractions of the semileptonic decay $\Lambda_b^0\to\Lambda_c^+\ell^-\overline{\nu}_\ell$ ($\ell$ stands for electron and muon) $\mathcal{B}r(\Lambda_b^0\to\Lambda_c^+\ell^-\overline{\nu}_\ell)=5.0\%$ and $\mathcal{B}r(\Lambda_b^0\to\Lambda_c^+\ell^-\overline{\nu}_\ell)/\mathcal{B}r(\Lambda_b^0\to\Lambda_c^+\pi^-)=16.6$, respectively \cite{DELPHI:2003qft,CDF:2008hqh}. With the ratio $\mathcal{R}(\Lambda_c^+)$ analyzed within the Standard Model, the early measurement of semileptonic decay of $\Lambda_b^0$ to $\Lambda_c^+$ with electron and muon final states gave the branching fraction of $\Lambda_b^0\to\Lambda_c^+\tau^-\overline{\nu}_\tau$ around two percent. Because of the poor experimental data of these processes, the newest observation and measurement also provided the material for testing the Standard Model by the ratio $\mathcal{R}(\Lambda_c^+)$.
   
     Theoretically, some methods have been used to calculate and analyze the processes $\Lambda_b^0\to\Lambda_c^+\ell^-\overline{\nu}_\ell$, such as QCD sum rules \cite{Dai:1996xv,Huang:2005mea,Azizi:2018axf}, light-front quark model \cite{Ke:2007tg,Zhu:2018jet,Ke:2019smy,Li:2021qod}, lattice QCD \cite{Detmold:2015aaa,Datta:2017aue}, heavy quark effective theory \cite{Huang:2005mea,Bernlochner:2018kxh,Han:2020sag}, relativistic quark model \cite{Faustov:2016pal}, covariant confined quark model \cite{Gutsche:2015mxa,Gutsche:2018nks}, Hypercentral constituent quark model \cite{Thakkar:2020vpv} and the analysis i searching for new physics \cite{Shivashankara:2015cta,Penalva:2019rgt} etc.. They calculated the branching fractions at some regions, and the results were consistent with experimental values. But with the analysis of tau lepton's final semileptonic decay, some references gave bigger results than the newest experiment or gave a large error. Besides, the Standard Model also gave a prediction of the ratio of branching fractions $\mathcal{R}(\Lambda_c^+)$, and the difference between experimental results and the values of the Standard Model predicted can be used to test the Standard Model or discover new physics beyond the Standard Model \cite{Mu:2019bin,Penalva:2020xup,Hu:2020axt,Penalva:2021gef,Azizi:2019tcn}. Based on these considerations, the reanalysis of the processes $\Lambda_b^0\to\Lambda_c^+\ell^-\overline{\nu}_\ell$ is needed and important.
     
     With the light-cone distribution amplitudes of $\Lambda_b$ developed from B-meson light-cone distribution amplitudes \cite{Ball:2008fw,Bell:2013tfa,Wang:2015ndk}, the light-cone sum rule approach is used in this article to calculate the correlation function of heavy baryon transition and study the properties of hadronic decay. Light-cone sum rule is a fruitful hybrid of the SVZ technique \cite{Shifman:1978bx,Shifman:1978by,Shifman:1978bw} and the theory of hard exclusive process \cite{Chernyak:1987nu,Colangelo:2000dp}. The basic idea of this approach is to expand the time-ordered products of hadrons and weak currents near the light-cone $x^2=0$. The method provides a powerful tool in the calculation of baryon transition form factors. With form factors calculated by the light-cone sum rule method, one can obtain the decay properties of hadrons. In this work, the method is used to calculate the form factors of the $\Lambda_b \to \Lambda_c$ transition. In the procedure to obtain the decay width and branching fractions of semileptonic decay $\Lambda_b^0 \to \Lambda_c^+ \ell^- \overline{\nu}_\ell (\ell =e, ~\mu, ~\tau)$ processes, we combine these form factors and the helicity form of differential decay width to obtain the decay properties of these processes. In order to compare the ratio of branching fractions $\mathcal{R}(\Lambda_c^+)=\mathcal{B}r(\Lambda_b^0 \to \tau^- \overline{\nu}_\tau)/\mathcal{B}r(\Lambda_b^0 \to \Lambda_c^+ \mu^- \overline{\nu}_\mu)$ predicted by the Standard Model and measured by experiments, $\mathcal{R}(\Lambda_c^+)$ is also given in this work.
    
     The article is arranged as follows: Sec:~\ref{sec:II} is the basic framework of the light-cone sum rule of the semileptonic decay $\Lambda_b^0 \to \Lambda_c^+ \ell^- \overline{\nu}_\ell$, in which the light-cone distribution amplitudes of bottom baryon $\Lambda_b$ are listed and the six form factors of weak transition $\Lambda_b \to \Lambda_c$ are given by the light-cone sum rule. Numerical analysis and the physical results of the $\Lambda_b \to \Lambda_c$ processes with two types of interpolating currents of $\Lambda_c$ states are given in Sec:~\ref{sec:III}. Sec:~\ref{sec:IV} are the conclusions and discussions.
    
    \section{Light-cone sum rules of $\Lambda_b \to \Lambda_c$ semileptonic decays}\label{sec:II}
   
   Before the theoretical analysis, one should convince oneself that the light-cone sum rule approach is valid for the physical process of bottom baryon for charm baryon decay. Similar to the light-cone sum rule analysis of $B\to D^{(*)}$ transition form factors, one expands the correlation function near the light-cone with the finite charm quark mass and with the light-cone dominance region of the interpolating charm baryon states momentum $p$ and momentum transfer $q$. With the same discussion as that in reference \cite{Faller:2008tr} and the applications in references \cite{Wang:2017jow,Gubernari:2018wyi,Aliev:2019ojc,Bordone:2019vic,Gao:2021sav,Gubernari:2022hrq}, the light-cone sum rule approach is used in this work  to calculate the $\Lambda_b^0\to\Lambda_c^+\ell^-\overline{\nu}_\ell$ processes. This approach employs a vacuum to an on-shell state correlation function and it is different from the conventional SVZ sum rule which is vacuum-to-vacuum. The other difference is that, in light-cone sum rule one expands the field operator product near the light-cone $x^2=0$ with a series of light-cone distribution amplitudes, while in the conventional SVZ sum rule one expands the field operator product near the position $x=0$ with a series of vacuum condensate operators. It gives us an easy way to calculate the form factors that appear in the three-point correlation function in the QCD sum rule method.
   
    \subsection{The basic framework of the semileptonic decay processes}
   
   The light-cone sum rule starts with the hadronic correlation function sandwiching the time-ordered product of the final hadronic current and weak current between vacuum and hadronic state, and the correlation function can be expressed as
    \begin{align}
   	T_\mu(p,q)=i\int d^4x e^{ip\cdot x} \langle 0|\mathcal{T}\{j_{\Lambda_c}(x),j_\mu(0)\}|\Lambda_b(p+q)\rangle,
   \end{align}
   where $p$ is the four-momentum of $\Lambda_c$ and $q$ is the momentum transfer. The interpolating current of heavy baryon $\Lambda_Q$ had been discussed in \cite{Shuryak:1981fza,Grozin:1992td,Ivanov:1996fj,Ivanov:1999pz,Zhang:2008pm}. There are two types of interpolating currents of $\Lambda_c$ in QCD sum rules, and they are given by
   \begin{gather}
   	j_{\Lambda_c}^1(x)=\epsilon_{ijk}[u^{iT}(x)C\gamma_5 d^j(x)]c^k(x), \label{current1}
   \end{gather}
   and
   \begin{gather}
   j_{\Lambda_c}^2(x)=\epsilon_{ijk}[u^{iT}(x)C\gamma_5\gamma_\nu d^j(x)]\gamma^\nu c^k(x). \label{current2}
   \end{gather}
   Both two choices of interpolating current can construct the light-cone sum rules of baryon decay. In the main text, we take the current type $j_{\Lambda_c}^1$ as an example to construct the light-cone sum rules of $\Lambda_b^0 \to \Lambda_c^+ \ell^- \overline{\nu}_{\ell}$ processes. The alternative interpolating current version $j_{\Lambda_c}^2$ can be obtained through the same procedures, and they are given in the appendix. The weak decay current of $b \to c$ is
   \begin{gather}
	j_\mu(0)=\overline{c}(0)\gamma_\mu(1-\gamma_5)b(0).
   \end{gather}

    In the light-cone sum rule approach, the correlation function can be derived at the hadronic and quark levels respectively. On the hadronic level, the correlation function can be parameterized by inserting a complete series of hadronic states with the same quantum number as the $\Lambda_c$ state, 
    \begin{equation}
    	\int d^4p\sum_i|\Lambda_c^i(p)\rangle\langle\Lambda_c^i(p)|=1.
    \end{equation}
    Where the index $i$ contains all states with the same quantum numbers of $\Lambda_c$.
    
    On the hadronic level, the correlation function can be represented as 
    \begin{align}
    T_\mu(p,q)&=\frac{\langle 0|j_{\Lambda_c}^i|\Lambda_c(p)\rangle \langle \Lambda_c(p)|j_{\mu}|\Lambda_b(p+q)}{M_{\Lambda_c}^2-p^2} \notag \\& +\frac{\langle 0|j_{\Lambda_c^*}^i|\Lambda_c^*(p)\rangle \langle \Lambda_c^*(p)|j_{\mu}|\Lambda_b(p+q)}{M_{\Lambda_c^*}^2-p^2} +\cdots. \label{CFH}
    \end{align}
     $i=1,2$ corresponds to two types of interpolating current $j_{\Lambda_c}^1$ and $j_{\Lambda_c}^2$. The correlation function on the hadronic level contains both the contributions of positive and negative parity interpolating states $\Lambda_c$ and $\Lambda_c^*$ baryons. $\langle 0|j_{\Lambda_c}|\Lambda_c(p)\rangle=f_{\Lambda_c}u_{\Lambda_c}(p)$ is the transition matrix element for the quantum number $(1/2)^+$ $\Lambda_c$ baryon, and $\langle 0|j_{\Lambda_c^*}|\Lambda_c^*(p)\rangle=f_{\Lambda_c^*}\gamma_5 u_{\Lambda_c^*}(p)$ is for the quantum number $(1/2)^-$ $\Lambda_c^*$ baryon. $f_{\Lambda_c}$ and $f_{\Lambda_c^*}$ are the decay constants, $u_{\Lambda_c}(p)$ and $u_{\Lambda_c^*}$ are the spinors of  $\Lambda_c$ and $\Lambda_c^*$, respectively. The ellipsis in the righthand side of the correlation function represents all the contributions of excited and continuum states with the same quantum number of $\Lambda_c$. The transition matrix element $\langle \Lambda_c(p)|j_\mu|\Lambda_b(p+q)\rangle$ can be parameterized by six form factors dependent on momentum transfer square $q^2$, and it's given by
    \begin{align}
    \langle \Lambda_c^{(*)}(p,s_1)|(V-A)_\mu|\Lambda_b(p+q,s_2)\rangle & = \overline{u}_{\Lambda_c^{(*)}}(p,s_1)[\gamma_\mu f_1^{(*)} (q^2)+i\frac{f_2^{(*)}(q^2)}{M_{\Lambda_b}}\sigma_{\mu\nu}q^{\nu}+\frac{f_3^{(*)}(q^2)}{M_{\Lambda_b}}q_\mu]u_{\Lambda_b}(p+q,s_2)\notag \\&
    -\overline{u}_{\Lambda_c^{(*)}}(p,s_1)[\gamma_\mu g_1^{(*)} (q^2)+i\frac{g_2^{(*)}(q^2)}{M_{\Lambda_b}}\sigma_{\mu\nu}q^{\nu}+\frac{g_3^{(*)}(q^2)}{M_{\Lambda_b}}q_\mu]\gamma_5 u_{\Lambda_b}(p+q,s_2). \label{transition matrix}
    \end{align}
 
  On the QCD level, by contracting the charm quark, the correlation function can be written as
  \begin{align}
  	T_\mu(p,q)=&i\int d^4 x e^{ip\cdot x}\langle 0|\mathcal{T}\{j_{\Lambda_c}(x),j_\mu(0)\}|\Lambda_b(p+q) \notag \\
  	=&i\int d^4xe^{ip\cdot x}(C\gamma_5)_{\alpha\beta}S_{\sigma\tau}(x)[\gamma_\mu(1-\gamma_5)]_{\tau\gamma} \langle 0|\epsilon_{ijk}u_\alpha^{iT}(x)d_\beta^j(x)b_\gamma^k(0)|\Lambda_b(p+q)\rangle, \label{CF}
    \end{align}
where $C$ is the charge conjugation matrix, and $S(x)$ is the free charm quark propagator.

   The light-cone distribution amplitude of $\Lambda_b$ baryon $\langle 0|\epsilon_{abc}u_\alpha^{aT}(x)d_\beta^b(x)b_\gamma^c(0)|\Lambda_b(p+q)\rangle$ has been investigated in \cite{Ball:2008fw} and used in the analysis of heavy baryon decay in \cite{Wang:2009hra,Wang:2009yma,Shi:2019fph}. We should notice that the light-cone distribution amplitudes are obtained in the heavy quark effective theory framework. And in reference \cite{Ball:2008fw}, the light-cone distribution amplitude $\langle 0|\epsilon_{abc}u^{aT}_\alpha(x)d^b_\beta(x)b^c_\gamma(0)|\Lambda_b(p+q)\rangle$ is represented by $\langle 0|\epsilon_{abc}u^{aT}_\alpha(x)d^b_\beta(x)b^c_\gamma(0)|\Lambda_b(v)\rangle$. Considering the heavy quark effective theory on both the hadronic and QCD level, the direct replacement $|\Lambda_b(p+q)\rangle \to |\Lambda_b(v)\rangle$ can be made safely. Therefore, the four-momentum of $\Lambda_b$ will be $p+q=M_{\Lambda_b}v$ with the on-shell condition. 
   
   At the quark level of the correlation function, the transition matrix element $\epsilon_{ijk}\langle 0|u_\alpha^{iT}(x)d_\beta^j(x)b_\gamma^k(0)|\Lambda_b(v)\rangle$ is
  
   \begin{align}
   \epsilon^{ijk}\langle 0|u^i_\alpha (t_1n)d^j_\beta(t_2n)h^k_{v\gamma}(0)|\Lambda_b(v)\rangle&= \frac{1}{8}f_{\Lambda_b}^{(2)}\Psi_2(t_1,t_2)(\slashed{\overline{n}}\gamma_5C)_{\alpha\beta}u_{\Lambda_b\gamma}(v)+\frac{1}{4}f_{\Lambda_b}^{(1)}\Psi_3^s(t_1,t_2)(\gamma_5C)_{\alpha\beta}u_{\Lambda_b\gamma}(v) \notag \\& 
   -\frac{1}{8}f_{\Lambda_b}^{(1)}\Psi_3^\sigma(t_1,t_2)(i\sigma_{\overline{n}n}\gamma_5C)_{\alpha\beta}u_{\Lambda_b\gamma}(v)+\frac{1}{8}f_{\Lambda_b}^{(2)}\Psi_4(t_1,t_2)(\slashed{n}\gamma_5C)_{\alpha\beta}u_{\Lambda_b\gamma}(v), \label{LCDA}
   \end{align}
   where $n_\mu$, $\overline{n}_\mu$, and $\sigma_{\overline{n}n}$ are
   \begin{gather*}
   	n_\mu=\frac{x_\mu}{v\cdot x},\quad \overline{n}_\mu=2v_\mu-\frac{x_\mu}{v\cdot x}, \quad  \sigma_{\overline{n}n}=\sigma_{\mu\nu}\overline{n}^{\mu}n^{\nu}.
   \end{gather*}
   The distribution amplitudes which have been given in \cite{Ball:2008fw} are
   \begin{equation}
   \Psi_i(t_1,t_2)=\int_{0}^{\infty}\omega d\omega\int_{0}^{1}due^{-i\omega(t_1u+t_2\overline{u})}\tilde{\psi}_i(\omega,u),
   \end{equation}
   with $\overline{u}=1-u$, $t_in=x_i$, and
   \begin{align}
   \tilde{\psi}_2(\omega,u)=&\omega^2u(1-u)[\frac{1}{\epsilon_0^4}e^{-\omega/\epsilon_0}+a_2C_2^{3/2}(2u-1)\frac{1}{\epsilon_1^4}e^{-\omega/\epsilon_1}],  \label{DA1} \\
   \tilde{\psi}_3^s(\omega,u)=&\frac{\omega}{2\epsilon_3^3}e^{-\omega/\epsilon_3}, \label{DA2} \\
   \tilde{\psi}_3^\sigma(\omega,u)=&\frac{\omega}{2\epsilon_3^3}(2u-1)e^{-\omega/\epsilon_3}, \label{DA3} \\
   \tilde{\psi}_4(\omega,u)=&5\mathcal{N}^{-1}\int_{\omega/2}^{s_0^{\Lambda_b}}dse^{-s/\tau}(s-\omega/2)^3. \label{DA4}
   \end{align}
   These parameters in the four light-cone distribution amplitudes are $\epsilon_0=200^{+130}_{-60}$ $\rm{MeV}, \epsilon_1=650^{+650}_{-300}$ $\rm{MeV}, \epsilon_3=230$ $\rm{MeV}$ and $a_2=0.333^{+0.250}_{-0.333}$. $C_2^{3/2}(2u-1)$ is the Gegenbauer polynomial. Other parameters and expressions and reliable regions, such as $s_0^{\Lambda_b}$ and $\tau$ can be found in reference \cite{Ball:2008fw}. The $\mathcal{N}$ in $\tilde{\psi}_4(\omega,u)$ is 
    \begin{align}
    \mathcal{N}=\int_0^{s_0} ds s^5 e^{-s/\tau}.
    \end{align}  
  
    Except for the above four light-cone distribution amplitudes, another definition will be useful in the following light-cone sum rules calculation in the $j_{\Lambda_c}^2$ interpolating current framework,
    
    \begin{align}
    \overline{\psi}_i(\omega,u)=\int_0^{\omega}d\tau \tau \tilde{\psi}_i(\tau, u).
    \end{align}
     where $\tilde{\psi}_i(\tau, u)$ correspond to light-cone distribution amplitudes (\ref{DA1}) - (\ref{DA4}).

   \subsection{Form factors of $\Lambda_b \to \Lambda_c$ transition}
  
   In order to calculate the decay widths and branching ratios of semileptonic decay $\Lambda_b^0 \to \Lambda_c^+ \ell^- \overline{\nu}_\ell$, the information of six form factors $f_i$ and $g_i$ $(i=1,2,3)$ should be known. It is known that the QCD sum rule method contains both the positive and negative parity contribution of ground states $\Lambda_c$ baryons with spin-1/2. To avoid the hidden uncertainty from the contribution of negative parity baryon $\Lambda_c^*$, the scheme developed in \cite{Jido:1996ia} which was used in QCD sum rules for nucleon and applied to light-cone sum rules for heavy baryon in \cite{Khodjamirian:2011jp} is taken into account. By substituting $\Lambda_c$ decay matrix element and $\Lambda_b \to \Lambda_c$ transition matrix elements (\ref{transition matrix}) of both positive and negative parity baryon $\Lambda_c$ and $\Lambda_c^*$ to Eq. (\ref{CFH}), and using the relations $\sum_s u_{\Lambda_c^{(*)}}(p,s)\overline{u}_{\Lambda_c^{(*)}}(p,s)=\slashed{p}+M_{\Lambda_c^{(*)}}$, the representation of correlation function on the hadronic level with the contribution both $\Lambda_c$ and $\Lambda_c^*$ can be expressed as 
    \begin{align}
   T_\mu(p,q)=& \frac{f_{\Lambda_c}}{M_{\Lambda_c}^2-p^2}\{2M_{\Lambda_b}f_1 v_\mu-[(M_{\Lambda_b} -M_{\Lambda_c})f_1  -\frac{M_{\Lambda_b}^2-M_{\Lambda_c}^2}{M_{\Lambda_b}}f_2]\gamma_\mu \notag \\&
   +[\frac{M_{\Lambda_b}+M_{\Lambda_c}}{M_{\Lambda_b}}(f_2  +f_3  -2f_1]q_\mu-2f_2(q^2)v_\mu \slashed{q}
    -[\frac{M_{\Lambda_b}+M_{\Lambda_c}}{M_{\Lambda_b}}f_2-f_1]\gamma_\mu\slashed{q} \notag \\&  +\frac{1}{M_{\Lambda_b}}[f_2-f_3]q_\mu\slashed{q} 
   -2M_{\Lambda_b}g_1(q^2)v_\mu\gamma_5-[(M_{\Lambda_b}+M_{\Lambda_c})g_1+\frac{M_{\Lambda_b}^2-M_{\Lambda_c}^2}{M_{\Lambda_b}}g_2]\gamma_\mu \gamma_5  \notag \\& 
   +[\frac{M_{\Lambda_b}-M_{\Lambda_c}}{M_{\Lambda_b}}(g_2 +g_3)+2g_1]q_\mu\gamma_5  +2g_2v_\mu\slashed{q}\gamma_5 
   -[g_1+\frac{M_{\Lambda_b}-M_{\Lambda_c}}{M_{\Lambda_b}}g_2]\gamma_\mu\slashed{q}\gamma_5\notag \\& -\frac{1}{M_{\Lambda_b}}[g_2(q^2)-g_3]q_\mu\slashed{q}\gamma_5\}u_{\Lambda_b}(v)  \notag \\&
  +\frac{f_{\Lambda_c^*}}{M_{\Lambda_c^*}^2-p^2}\{-2M_{\Lambda_b}f_1^*v_\mu  +[(M_{\Lambda_b}+M_{\Lambda_c^*})f_1^* +\frac{M_{\Lambda_b}^2-M_{\Lambda_c}^2}{M_{\Lambda_b}}f_2^*]\gamma_\mu  \notag \\& +[\frac{M_{\Lambda_b}-M_{\Lambda_c^*}}{M_{\Lambda_b}}(f_2^*+f_3^*)+2f_1^*]q_\mu-2f_2^*v_\mu\slashed{q}+[\frac{M_{\Lambda_c^*}-M_{\Lambda_b}}{M_{\Lambda_b}}f_2^*-f_1^*]\gamma_\mu\slashed{q} \notag \\& +\frac{1}{M_{\Lambda_b}}(f_2^*-f_3^*)q_\mu\slashed{q}
   +2M_{\Lambda_b}g_1^*v_\mu\gamma_5+[(M_{\Lambda_b} -M_{\Lambda_c^*})g_1^*+\frac{M_{\Lambda_c}^2-M_{\Lambda_b}^2}{M_{\Lambda_b}}g_2^*]\gamma_\mu\gamma_5  \notag \\& +[\frac{M_{\Lambda_b}+M_{\Lambda_c^*}}{M_{\Lambda_b}}(g_2^*+g_3^*)-2g_1^*]q_\mu\gamma_5   +2g_2^*v_\mu\slashed{q}\gamma_5 +[g_1^*-\frac{M_{\Lambda_b}+M_{\Lambda_c^*}}{M_{\Lambda_b}}g_2^*]\gamma_\mu\slashed{q}\gamma_5 \notag \\& -\frac{1}{M_{\Lambda_b}}(g_2^*-g_3^*)q_\mu\slashed{q}\gamma_5
   \}u_{\Lambda_b}(v)+\cdots.
    \end{align}
    
   Substituting the light-cone distribution amplitudes of $\Lambda_b$ Eq. (\ref{LCDA}) into the correlation function Eq. (\ref{CF}), the expression on the QCD level is obtained as
    \begin{align}
    T_\mu(p,q)=&-\int_{0}^{\infty}\omega d\omega \int_{0}^{1}du\tilde{\psi}_3^s(\omega,u)\frac{f_{\Lambda_b}^{(1)}}{(p-\omega v)^2-m_c^2}[2(M_{\Lambda_b}-\omega)v_\mu+\gamma_\mu \slashed{q}-2q_\mu  \notag \\& -(M_{\Lambda_b}-\omega-m_c)\gamma_\mu 
    -2(M_{\Lambda_b}-\omega)v_\mu\gamma_5-\gamma_\mu\slashed{q}\gamma_5+2q_\mu\gamma_5 \notag \\& -(M_{\Lambda_b}-\omega+m_c)\gamma_\mu\gamma_5]u_{\Lambda_b}(v).
    \end{align}
    
     Comparing with the same coeffecients of Lorentz structures $\Gamma=\{v_\mu, \gamma_\mu, q_\mu, v_\mu\slashed{q}, \gamma_\mu \slashed{q}, q_\mu\slashed{q}\}$ and $\Gamma \gamma_5$ on both the hadronic and QCD levels, one can substract and eliminate the contributions of negative parity baryon $\Lambda_c^*(\frac{1}{2})^-$ in the light-cone sum rules by solving the linear equations of form factors. Thereafter,
    one can make a standard light-cone sum rule calculation, and a Borel transformation to suppress both the higher twist and continuum contributions
    \begin{align}
    	\int_{0}^{\infty}d\sigma\frac{\rho_i(\sigma)}{(p-\omega v)^2-m_c^2} \to -\int_{0}^{\sigma_0}d\sigma\frac{\rho_i(\sigma)e^{-s/M_B^2}}{\overline{\sigma}},
    \end{align}
    where $\sigma=\omega/M_{\Lambda_b}$ and $\overline{\sigma}=1-\sigma$. We should also notice that in the heavy quark effective theory the decay constant of $\Lambda_b$ and $\Lambda_c$ has no difference, which has been discussed in reference \cite{Huang:2005mea}. With the quark-hadron duality assumption, after making the light-cone sum rule calculation procedure, and excluding the negative paity $\Lambda_c^*$ contribution, one has the relations of form factors $f_1(q^2)=g_1(q^2)$ and $f_2(q^2)=f_3(q^2)=g_2(q^2)=g_3(q^2)$. Form factors $f_1(q^2)$ and $f_2(q^2)$ can be expressed as
    \begin{align}
    f_1(q^2)=&\int_0^1 du \int_0^{\sigma_0} d\sigma \frac{\sigma M_{\Lambda_b}^2(m_c-\sigma M_{\Lambda_c}+M_{\Lambda_c^*})}{\overline{\sigma}(M_{\Lambda_c}+M_{\Lambda_c^*})} \psi_3^s(\omega,u)e^{(M_{\Lambda_c}^2-s)/M_B^2}, \\
    f_2(q^2)=&-\int_0^1 du \int_0^{\sigma_0} d\sigma \frac{\sigma^2 M_{\Lambda_b}^3\psi_3^s(\omega,u)}{\overline{\sigma}(M_{\Lambda_c}+M_{\Lambda_c^*})}e^{(M_{\Lambda_c}^2-s)/M_B^2}. \label{form factors}
    \end{align}
    The parameter $M_B$ is the Borel mass and the extra parameters introduced in the above equations are defined as
        \begin{gather}
         s=\sigma M_{\Lambda_b}^2+\frac{m_c^2-\sigma q^2}{\overline{\sigma}},
        \\ 
        \sigma_0=\frac{(s_0+M_{\Lambda_b}^2-q^2)}{2M_{\Lambda_b}^2}-\frac{\sqrt{(s_0+M_{\Lambda_b}^2-q^2)^2-4M_{\Lambda_b}^2(s_0-m_c^2)}}{2M_{\Lambda_b}^2}, 
        \end{gather}
        here $s_0$ is the threshold of $\Lambda_c$ baryon. One can notice that the mass of $\Lambda_c^*$ enters the sum rules through the practice of eliminating the contribution of negative parity baryon $\Lambda_c^*$.
       
        \section{Numerical analysis}\label{sec:III}
      
          The input parameters in the numerical analysis, such as the mass of heavy baryons and leptons, are adopted from PDG \cite{Zyla:2020zbs} and listed in Table \ref{table 1}. The heavy charm quark mass $m_c$ adopted in this work is the pole mass. Since the pole mass is a gauge-invariant, infrared finite, and renormalization-scheme-independent quantity \cite{Tarrach:1980up}. The pole mass of charm quark $m_c=(1.67\pm 0.07)$ $\rm{GeV}$ corresponds to the $\overline{MS}$ mass $\overline{m}_c=(1.27\pm 0.02)$ $\rm{GeV}$ in PDG, and the relations between $\overline{MS}$ and pole mass can be found in \cite{Zyla:2020zbs, Narison:2002woh}.
          \begin{table}[h]
          	\centering
          	\caption{Masses of heavy baryons, quarks and leptons.} \label{table 1}
          	\begin{tabular}{cc}\hline
          		Parameters&PDG Values \\ \hline
          		$M_{\Lambda_b}$&$5.6196$ $\rm{GeV}$ \\
          		$M_{\Lambda_c}$&$2.28646$ $\rm{GeV}$ \\
          		$M_{\Lambda_c^*}$&$2.59225$ $\rm{GeV}$ \\
          		$m_c$&$(1.67\pm 0.07)$ $\rm{GeV}$ \cite{Gutsche:2015mxa, Zyla:2020zbs}\\
          		$m_e$&$0.511$ $\rm{MeV}$ \\
          	    $m_\mu$&$105.66$ $\rm{MeV}$ \\
          	    $m_\tau$&$1.77686$ $\rm{GeV}$ \\ \hline
          	\end{tabular}
          \end{table}
         
          \begin{figure*}[!]
          	\includegraphics[width=0.45\textwidth]{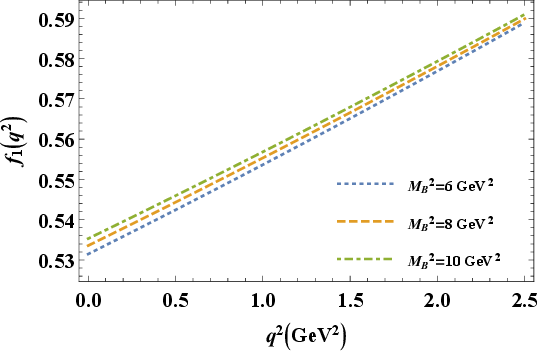}
          	\qquad
          	\includegraphics[width=0.45\textwidth]{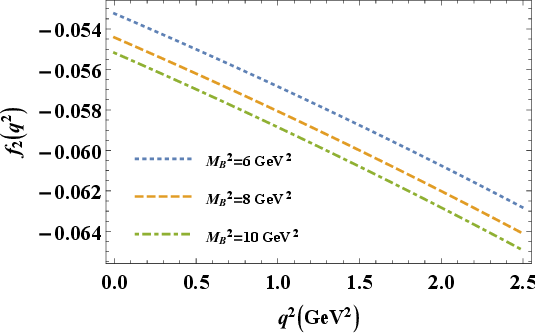}
          	\caption{\label{fig1}The form factors of $f_1(q^2)$ and $f_2(q^2)$ depend on the momentum transfer $q^2$ with the values $m_c=1.67~\rm{GeV},~s_0=(M_{\Lambda_c}+0.4)^2~\rm{GeV^2}$ at $M_B^2=6$ $\rm{GeV^2}$, $M_B^2=8$ $\rm{GeV^2}$ and $M_B^2=10$ $\rm{GeV^2}$ with each form factor of the current type $j_{\Lambda_c}^1$.}
          \end{figure*}
          
         With these parameters and equations of form factors, one can calculate the semileptonic decay process. One considers the threshold $s_0$ of $\Lambda_c$ set to be larger than the ground states baryon $\Lambda_c$ and $\Lambda_c^*$ mass, but lower than the excited states baryon $\Lambda_c$ mass. So, we set $s_0=(M_{\Lambda_c}+\Delta)^2$, $\Delta=(0.40\pm 0.05)$ $\rm{GeV}$. The Borel parameter $M_B$ is set to the interval where the form factors have a stable region with $M_B^2$, and should also suppress the contribution of excited and continuum states of final baryon $\Lambda_c$ and the higher twist of the initial baryon $\Lambda_b$ light-cone distribution amplitudes. With these requirements, the Borel parameter $M_B$ is chosen as $M_B^2=(8\pm 2)$ $\rm{GeV^2}$. The form factors relying on the momentum transfer square $q^2$ are calculated with the current $j_{\Lambda_c}^1$ and $j_{\Lambda_c}^2$ respectively based on the above considerations. With the threshold $(M_{\Lambda_c}+0.35)^2~\rm{GeV^2}  \le s_0 \le (M_{\Lambda_c}+0.45)^2~\rm{GeV}^2$, and Borel parameters $M_B^2$ $6~\rm{GeV^2} \le M_B^2 \le 8~\rm{GeV^2}$ respectively, form factors $F_i(q^2)$ with the interpolating current $j_{\Lambda_c}^1$ at $q^2=0$ $\rm{GeV^2}$ are $f_1(0)=g_1(0)=0.534^{+0.060}_{-0.074}$, $f_2(0)=f_3(0)=g_2(0)=g_3(0)=-0.054^{+0.013}_{-0.011}$, the errors come from the uncertainties of input parameters and the regions of Borel parameters $M_B$ and threshold $s_0$.
        
        As discussed in reference \cite{Huang:2004vf} and our numerical analysis, the light-cone sum rule is not applicable to the whole physical region $m_\ell^2\le q^2 \le (M_{\Lambda_b}-M_{\Lambda_c})^2$. In this work, the only applicable region of $q^2$ is $0$ $\rm{GeV}^2\le q^2 \le 2.5$ $\rm{GeV}^2$, as shown in Fig. \ref{fig1}. For the whole physical regions, we should extrapolate the form factors obtained by the light-cone sum rule from the light-cone sum rule applicable region to the whole physical region through a fitting formula. In this work, the following general ``z-expansion" formula is used to fit the form factors and extrapolate them to the whole physical region \cite{Bourrely:2008za}.
         
         \begin{align}
         F_i(q^2)=\frac{F_i(0)}{1-\frac{q^2}{M_{B_c}^2}}\{1+b_1[z(q^2,t_0)-z(0,t_0)]+b_2[z(q^2,t_0)^2-z(0,t_0)^2]\},
         \end{align}
        where 
         \begin{align}
         z(q^2,t_0)=\frac{\sqrt{t_+-q^2}-\sqrt{t_+-t_0}}{\sqrt{t_+-q^2}+\sqrt{t_+-t_0}},
         \end{align}
         and 
         \begin{gather}
         t_+=(M_{\Lambda_b}+M_{\Lambda_c})^2, \notag \\ t_0=(M_{\Lambda_b}+M_{\Lambda_c})\cdot (\sqrt{M_{\Lambda_b}}-\sqrt{M_{\Lambda_c}})^2.
         \end{gather}
         $M_{B_c}=6.27447~\rm{GeV}$ is the mass of $B_c$ meson \cite{Zyla:2020zbs}. $F_i(0)$ is the form factor at $q^2=0~\rm{GeV^2}$.
         
         With the fitting formula and the data of form factors at $q^2=0$ $\rm{GeV^2}$ and the interval $0\le q^2\le 2.5$ $\rm{GeV^2}$, the fitting parameters of form factors $F_i(q^2)$ with the interpolating current type $j_{\Lambda_c}^1$ can be obtained and are listed in Table \ref{table 2}.
         
          \begin{table}[h]
          	\centering
          	\caption{Fitting parameters of form factors $F_i(q^2)$ with the interpolating current type $j_{\Lambda_c}^1$, with the errors coming from the uncertainties of input parameters, Borel parameters $M_B$ and threshold $s_0$.} \label{table 2}	
          	\begin{tabular}{cccc}\hline
          	$F_i(q^2)$&$F_i(0)$&$b_1$&$b_2$ \\ \hline
          	$f_1(q^2)$&$0.534^{+0.060}_{-0.074}$&$-2.883^{+0.779}_{-1.619}$&$-15.991^{+9.571}_{-3.108}$ \\ 
          	$f_2(q^2)$&$-0.054^{+0.013}_{-0.011}$&$-11.474^{+2.262}_{-2.369}$&$34.934^{+21.608}_{-12.878}$ \\
            \hline          		
          	\end{tabular}
          \end{table}
        
             \begin{figure*}[!]
             	\includegraphics[width=0.3\textwidth]{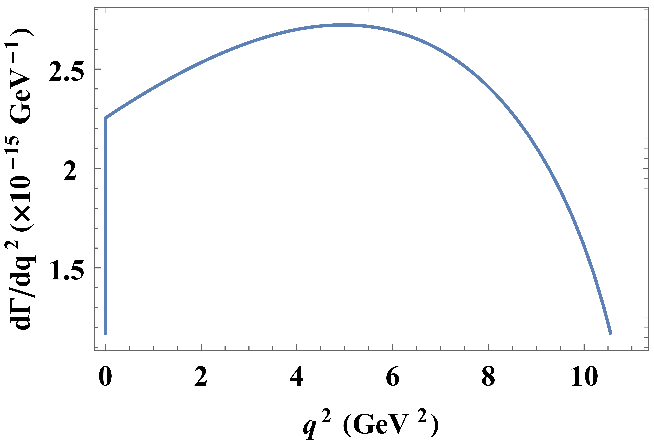}
             	\qquad
             	\includegraphics[width=0.3\textwidth]{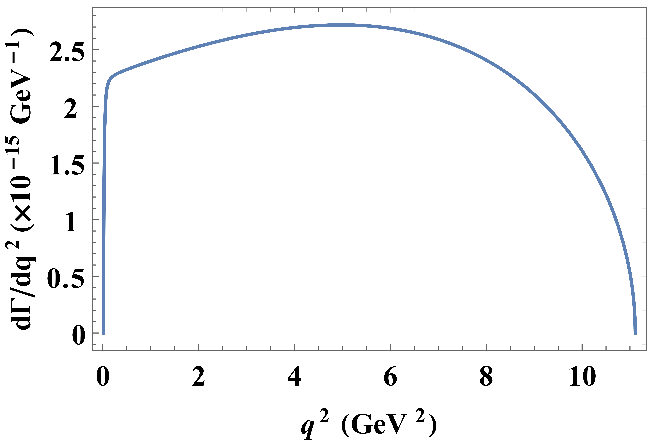}
             	\qquad
             	\includegraphics[width=0.3\textwidth]{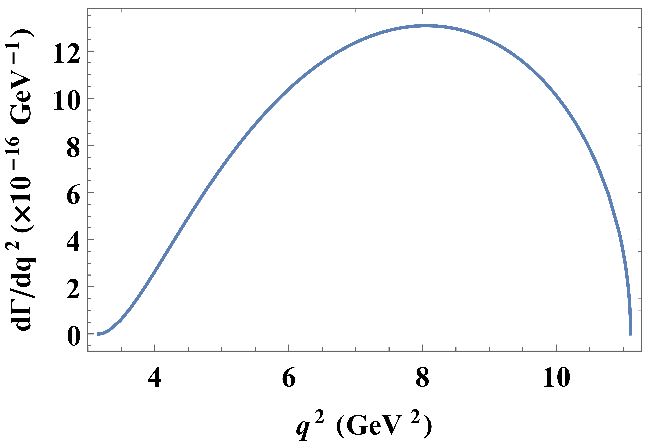}
             	\caption{\label{fig2} Differential decay widths of the current type $j_{\Lambda_c}^1$ of lepton final state with electron (left), muon (middle), and tau (right) respectively. The figures are plotted at the central values $m_c=1.67~\rm{GeV}$, $s_0=(M_{\Lambda_c}+0.4)^2~\rm{GeV^2}$ and $M_B^2=8~\rm{GeV^2}$.} 
             \end{figure*}    
            
      To obtain the branching fractions of semileptonic decays of $\Lambda_b^0$ to $\Lambda_c^+$, one should know the lifetime or partial decay width of  $\Lambda_b^0$.  The lifetime of $\Lambda_b$ baryon is adopted from PDG, with $\tau_{\Lambda_b}=(1.471\pm 0.009)$ $\rm{ps}$. The helicity amplitude forms of semileptonic decay widths \cite{Bialas:1992ny} are used in these processes. One has the following equation of differential decay width written as two polarized decay widths
      \begin{gather}
      \frac{d\Gamma}{dq^2}=\frac{d\Gamma_L}{dq^2}+\frac{d\Gamma_T}{dq^2},
      \end{gather}
      where $\Gamma_L$ and $\Gamma_T$ are longitudinally and transversely polarized decay widths, respectively. The total decay width is
      \begin{equation}
      \Gamma=\int_{m_l^2}^{(M_{\Lambda_b}-M_{\Lambda_c})^2}dq^2\frac{d\Gamma}{dq^2},
      \end{equation}
      where
      \begin{align}
      \frac{d\Gamma_L}{dq^2}&=\frac{G_F^2|V_{cb}|^2 q^2 p(1-\hat{m}_l^2)^2}{384\pi^3 M_{\Lambda_b}^2}[(2+\hat{m}_l^2)(|H_{\frac{1}{2},0}|^2 \notag \\&+|H_{-\frac{1}{2},0}|^2)+3\hat{m}_l^2(H_{\frac{1}{2},t}|^2+H_{-\frac{1}{2},t}|^2)],  \\
      \frac{d\Gamma_T}{dq^2}&=\frac{G_F^2|V_{cb}|^2 q^2 p(1-\hat{m}_l^2)^2(2+\hat{m}_l^2)}{384\pi^3 M_{\Lambda_b}^2}(|H_{\frac{1}{2},1}|^2 \notag \\& +|H_{-\frac{1}{2},-1}|^2).
      \end{align}
      
      In the above equations, $p=\sqrt{Q_+Q_-}/2M_{\Lambda_b}, Q_\pm=(M_{\Lambda_b}\pm M_{\Lambda_c})^2-q^2$, and $\hat{m}_l\equiv m_l/\sqrt{q^2}$, in which $m_l$ is the mass of lepton ($e,~\mu$ or $\tau$). The relevant expressions of the helicity connected with form factors are given by
      \begin{align}
      H_{\frac{1}{2},0}^V=&-i\frac{\sqrt{Q_-}}{\sqrt{q^2}}[(M_{\Lambda_b}+M_{\Lambda_c})f_1(q^2)-\frac{q^2}{M_{\Lambda_b}}f_2(q^2)], \\
      H_{\frac{1}{2},1}^V=&i\sqrt{2Q_-}[-f_1(q^2)+\frac{M_{\Lambda_b}+M_{\Lambda_c}}{M_{\Lambda_b}}f_2(q^2)],  \\
      H_{\frac{1}{2},0}^A=&-i\frac{\sqrt{Q_+}}{\sqrt{q^2}}[(M_{\Lambda_b}-M_{\Lambda_c})g_1(q^2)+\frac{q^2}{M_{\Lambda_b}}g_2(q^2)], \\
      H_{\frac{1}{2},1}^A=&i\sqrt{2Q_+}[-g_1(q^2)-\frac{M_{\Lambda_b}-M_{\Lambda_c}}{M_{\Lambda_b}}g_2(q^2)], \\
      H_{\frac{1}{2},t}^V=&-i\frac{\sqrt{Q_+}}{q^2}[(M_{\Lambda_b}-M_{\Lambda_c})f_1(q^2)+\frac{q^2}{M_{\Lambda_b}}f_3(q^2)], \\
      H_{\frac{1}{2},t}^A=&-i\frac{Q_-}{\sqrt{q^2}}[(M_{\Lambda_b}+M_{\Lambda_c})g_1(q^2)-\frac{q^2}{M_{\Lambda_b}}g_3(q^2)].
      \end{align}
      The negative helicity amplitudes can be obtained through the positive helicity amplitudes as
      \begin{gather}
      H_{-\lambda,-\lambda_W}^V=H_{\lambda,\lambda_W}^V,   \notag \\     H_{-\lambda,-\lambda_W}^A=-H_{\lambda,\lambda_W}^A.
      \end{gather}
      $\lambda$ and $\lambda_W$ are the polarizations of the final $\Lambda_c$ baryon and W-Boson, respectively.
      
      With the $V-A$ current, the total helicity amplitudes are expressed as 
      \begin{gather}
      H_{\lambda,\lambda_W}=H_{\lambda,\lambda_W}^V-H_{\lambda,\lambda_W}^A.
      \end{gather}

        \begin{table}[h]
        	\centering
        	\caption{The decay widths $\Gamma(\Lambda_b^0 \to \Lambda_c^+ \ell^- \overline{\nu}_\ell)$, branching fractions $\mathcal{B}r(\Lambda_b^0\to\Lambda_c^+\ell^-\overline{\nu}_\ell)$ and ratios $\Gamma_L/\Gamma_T$ of current type $j_{\Lambda_c}^1$.} \label{table 3}
        	\begin{tabular}{cccc} \hline
        		$\Lambda_b^0 \to \Lambda_c^+ \ell^- \overline{\nu}_\ell$&$\Gamma (\times 10^{-14} \rm{GeV})$&$\mathcal{B}r (\%)$&$\Gamma_L/\Gamma_T$ \\ \hline
        		$\Lambda_b^0 \to \Lambda_c^+ e^- \overline{\nu}_e$&$2.60_{-0.54}^{+0.52} $&$5.81^{+1.16}_{-1.21}$&$1.796^{+0.033}_{-0.061}$  \\ 
        		$\Lambda_b^0 \to \Lambda_c^+ \mu^- \overline{\nu}_\mu$&$2.59^{+0.52}_{-0.54}$&$5.79^{+1.15}_{-1.20}$&$1.794^{+0.033}_{-0.061}$  \\ 
        		$\Lambda_b^0 \to \Lambda_c^+ \tau^- \overline{\nu}_\tau$&$0.71^{+0.13}_{-0.13}$&$1.59^{+0.28}_{-0.29}$&$1.478^{+0.013}_{-0.024}$  \\ \hline
        	\end{tabular}		
        \end{table} 
          
      One substitutes the form factors of weak decay $\Lambda_b \to \Lambda_c$ into the helicity form of decay widths, uses these basic parameters provided by PDG, and considers the CKM matrix element $|V_{cb}|=(41.0\pm 1.4)\times 10^{-3}$ \cite{Zyla:2020zbs}, then integrates the momentum transfer square $q^2$ on the whole physical region. Therefore, the information on decay widths $\Gamma(\Lambda_b^0\to\Lambda_c^+\ell^-\overline{\nu}_\ell)$ and the ratios $\Gamma_L/\Gamma_T$ can be known. The absolute branching ratios $\mathcal{B}r(\Lambda_b^0\to\Lambda_c^+\ell^-\overline{\nu}_\ell)$ can be calculated with the lifetime of $\Lambda_b$ baryon and results are listed in Table \ref{table 3}. To compare our results with other works, the unit second inverse is used and listed in Table \ref{table 4}. The pictures of differential decay widths of $\Lambda_b^0 \to \Lambda_c^+ \ell^- \overline{\nu}_\ell$ with current $j_{\Lambda_c}^1$ at central values of parameters are shown in Fig. \ref{fig2}.

      \begin{table}[h]
      	\centering
      	\caption{The decay widths $\Gamma(\Lambda_b^0 \to \Lambda_c^+ \ell^- \overline{\nu}_\ell)$, branching fractions $\mathcal{B}r(\Lambda_b^0\to\Lambda_c^+\ell^-\overline{\nu}_\ell)$ and ratios $\Gamma_L/\Gamma_T$ of current type $j_{\Lambda_c}^2$.} \label{table 32}
      	\begin{tabular}{cccc} \hline
      		$\Lambda_b^0 \to \Lambda_c^+ \ell^- \overline{\nu}_\ell$&$\Gamma (\times 10^{-14} \rm{GeV})$&$\mathcal{B}r(\%)$&$\Gamma_L/\Gamma_T$ \\ \hline
      		$\Lambda_b^0 \to \Lambda_c^+ e^- \overline{\nu}_e$&$3.03^{+1.30}_{-2.03} $&$6.76^{+2.90}_{-4.53}$&$2.073^{+0.165}_{-0.433}$  \\ 
      		$\Lambda_b^0 \to \Lambda_c^+ \mu^- \overline{\nu}_\mu$&$3.01^{+1.29}_{-2.02}$&$6.73^{+2.88}_{-4.51}$&$2.070^{+0.164}_{-0.430}$  \\ 
      		$\Lambda_b^0 \to \Lambda_c^+ \tau^- \overline{\nu}_\tau$&$0.72^{+0.22}_{-0.41}$&$1.61^{+0.49}_{-0.92}$&$1.599^{+0.063}_{-0.153}$  \\ \hline
      	\end{tabular}		
      \end{table}
     
     Turn to another case of $\Lambda_c$ interpolating current $j_{\Lambda_c}^2(x)=\epsilon_{ijk}[u^{iT}(x)C\gamma_5\gamma_\nu d^j(x)]\gamma^\nu c^k(x)$. By using the same procedures as those for the current type $j_{\Lambda_c}^1$ with the same parameters as those in the above analysis,and then obtains the same relations of form factors  $f_1(q^2)=g_1(q^2)$, $f_2(q^2)=f_3(q^2)=g_2(q^2)=g_3(q^2)$ as well as that in the case of $j_{\Lambda_c}^1$. Setting the same threshold $s_0$ and the Borel mass $M_B^2$ as in the case of $j_{\Lambda_c}^1$, form factors with the interpolating current type $j_{\Lambda_c}^2$ at $q^2=0$ $\rm{GeV}^2$ give the values $f_1(0)=g_1(0)=0.644^{+0.175}_{-0.352}$, $f_2(0)=f_3(0)=g_2(0)=g_3(0)=-0.100^{+0.061}_{-0.039}$, where the uncertainties contain the errors of input parameters, thresholds, and Borel parameters. Using the helicity amplitudes form of decay width, the decay widths $\Gamma(\Lambda_b^0\to\Lambda_c^+\ell^-\overline{\nu}_\ell)$, branching fractions $\mathcal{B}r(\Lambda_b^0\to\Lambda_c^+\ell^-\overline{\nu}_\ell)$, and the ratios $\Gamma_L/\Gamma_T$ within the interpolating current $j_{\Lambda_c}^2$ can be obtained and they are shown in Table \ref{table 32}. 
         
      \begin{table*}[htbp]
      	\centering
      	\caption{The decay widths and branching fractions compared with other models. The first and second values in the $\Gamma$ and $\mathcal{B}r$ columns stand for the $\Lambda_b^0 \to \Lambda_c^+ e^- \overline{\nu}_e$ and $\Lambda_b^0 \to \Lambda_c^+ \mu^- \overline{\nu}_\nu$, respectively. All the data listed are the central values} \label{table 4}
      	\begin{tabular}{ccccc} \hline
      		\multirow{2}*{References}&\multicolumn{2}{c}{Decay widths $\Gamma(\times 10^{10} \rm{s}^{-1})$}&\multicolumn{2}{c}{Branching fractions ($\times 10^{-2}$)} \\  &$\Gamma(\Lambda_b^0\to\Lambda_c^+\ell^-\overline{\nu}_\ell)$&$\Gamma(\Lambda_b^0\to\Lambda_c^+\tau^-\overline{\nu}_\tau)$&$\mathcal{B}r(\Lambda_b^0\to\Lambda_c^+\ell^-\overline{\nu}_\ell)$&$\mathcal{B}r(\Lambda_b^0\to\tau^-\overline{\nu}_\tau)$    \\ \hline
      		 LHCb   \cite{LHCb:2022piu}&-&-&-&1.50 \\
      		 DELPHI \cite{DELPHI:2003qft}&-&-&5.0&- \\
      		 CDF    \cite{CDF:2008hqh}&-&-&7.3&2.0 \\
      		        \cite{Huang:2005mea}&5.4&-&-&- \\
      		        \cite{Azizi:2018axf}&3.52&1.12&6.04&1.87 \\
      		        \cite{Ke:2007tg}&-&-&6.2, 6.3&- \\
      		        \cite{Zhu:2018jet}&-&-&5.59, 5.57&1.54 \\
      		     	\cite{Ke:2019smy}&5.0, 7.7&-&-&- \\
      		    	\cite{Li:2021qod}&-&-&6.47, 6.45&1.97 \\
      		        \cite{Detmold:2015aaa}&3.61&1.2&-&- \\
      		        \cite{Faustov:2016pal}&4.42,4.41&1.39&6.48, 6.46&2.03 \\
      	            \cite{Gutsche:2015mxa}&-&-&6.9&2.0 \\
      	            \cite{Thakkar:2020vpv}&4.11&-&6.04&- \\ 
      		        \cite{Mu:2019bin}&-&-&5.34&1.78 \\
      		        \cite{Singleton:1990ye}&5.9&-&-&- \\
      		        \cite{Cheng:1995fe}&5.1&-&-&- \\
      		        \cite{Ivanov:1996fj}&5.39&-&-&- \\
      		        \cite{Ivanov:1998ya}&6.09&-&-&- \\
      		        \cite{MarquesdeCarvalho:1999bqs}&5.01, 7.61, 2.73&-&-& \\
      		        \cite{Cardarelli:1998tq}&-&-&6.3&- \\
      		        \cite{Albertus:2004wj}&5.82&-&-&- \\
      		        \cite{Ebert:2006rp}&5.02, 5.64&-&6.2, 6.9&- \\
      		        \cite{Zhao:2020mod}&4.50&-&6.61&- \\ 
       		       This work $j_{\Lambda_c}^1$&3.95, 3.94&1.08&5.81, 5.79&1.59 \\
       		       This work $j_{\Lambda_c}^2$&4.60, 4.57&1.09&6.76, 6.73&1.61\\
      		  \hline
      	\end{tabular}
      \end{table*}

 \begin{table*}[htbp]
 	\centering
 	\caption{Ratios of branching fractions $\mathcal{R}(\Lambda_c^+)$ compared with others' work with the central values} \label{table 5}
 	\begin{tabular}{cccccccccccccc} \hline 
 		References&Exp. \cite{LHCb:2022piu}&\cite{Azizi:2018axf}&\cite{Zhu:2018jet}&\cite{Detmold:2015aaa}&\cite{Bernlochner:2018kxh,Bernlochner:2018bfn}&\cite{Faustov:2016pal}&\cite{Gutsche:2015mxa}&\cite{Shivashankara:2015cta}&\cite{Mu:2019bin}&\cite{Penalva:2021gef}&\cite{Becirevic:2020nmb}&$j_{\Lambda_c}^1$&$j_{\Lambda_c}^2$  \\ \hline 
 		$\mathcal{R}(\Lambda_c^+)$&0.242&0.31&0.28&0.333&0.324&0.313&0.294&0.29&0.33&0.317&0.332&0.274&0.239 \\ \hline
 	\end{tabular}
 \end{table*}
 
      The ratio of semileptonic decay branching fractions with tau lepton final states and light lepton final states $\mathcal{R}(H_c)$ ($H$ stands for hadron), in the early experimental measurement was only reported in the $B\to D(D^*)$ weak decay. BaBar Collaboration gave their evidence $\mathcal{R}(D)=(0.440\pm0.058)$ and $\mathcal{R}(D^*)=(0.332\pm0.024)$ \cite{BaBar:2012obs}. Meanwhile the Standard Model gave the prediction $\mathcal{R}(D)=(0.297\pm0.017)$ and $\mathcal{R}(D^*)=(0.252\pm0.003)$ \cite{Fajfer:2012vx}. It shows that there is a larger value in the experiment than the Standard Model prediction, which is $\mathcal{R}(D^*)$ puzzle. The recent experiment LHCb reported $\mathcal{R}(\Lambda_c^+)=(0.242\pm0.026\pm0.040\pm0.059)$, with the last uncertainty from the measurement of branching fraction uncertainty of the channel $\Lambda_b^0\to\Lambda_c^+\mu^-\overline{\nu}_\mu$, and what they reported in their experiment agrees with the prediction of the Standard Model. In this work, the values of $\mathcal{R}(\Lambda_c^+)$ that we calculate are $\mathcal{R}(\Lambda_c^+)=0.274^{+0.009}_{-0.005}$ with the $\Lambda_c$ interpolating current type $j_{\Lambda_c}^1$, and $\mathcal{R}(\Lambda_c^+)=0.239^{+0.070}_{-0.021}$ with the $\Lambda_c$ interpolating current type $j_{\Lambda_c}^2$, which are also consistent with the recent experimental report and the prediction of the Standard Model. The comparison of the results in our works with other models and experiments' is displayed in Table \ref{table 5}.

     \section{Conclusions and Discussions}\label{sec:IV}
     
     In the light-cone sum rules approach, the six transition form factors of $\Lambda_b\to\Lambda_c$ weak decay are calculated with the $\Lambda_b$ baryon light-cone distribution amplitudes. In the calculations, we use two types of $\Lambda_c$ interpolating currents $j_{\Lambda_c}^1$ and $j_{\Lambda_c}^2$, and give the relations of these form factors $f_1(q^2)=g_1(q^2)$, $f_2(q^2)=f_3(q^2)=g_2(q^2)=g_3(q^2)$ for both types of $\Lambda_c$ interpolating current $j_{\Lambda_c}^1$ and $j_{\Lambda_c}^2$. The values of six form factors at $q^2=0$ $\rm{GeV^2}$ obtained by the light-cone sum rule with $\Lambda_b$ baryon distribution amplitudes give that $f_1(0)=g_1(0)=0.533^{+0.060}_{-0.074}$, $f_2(0)=f_3(0)=g_2(0)=g_3(0)=-0.054^{+0.013}_{-0.011}$ with interpolating current $j_{\Lambda_c}^1$ and $f_1(0)=g_1(0)=0.644^{+0.175}_{-0.352}$, $f_2(0)=f_3(0)=g_2(0)=g_3(0)=-0.100^{+0.061}_{-0.039}$ with interpolating current $j_{\Lambda_c}^2$, which are all in accordance with the results in other methods and models' and have similar results to heavy quark effective theory analysis \cite{Bernlochner:2018kxh}. When these form factors are combined with the helicity amplitude to calculate the differential decay widths and branching fractions of heavy baryon semileptonic decay $\Lambda_b^0\to\Lambda_c^+\ell^-\overline{\nu}_\ell$, the results in our work is in agreement with those in other theoretical models and experimental reports.
     
     The uncertainties of form factors and decay rates with interpolating current $j_{\Lambda_c}^1$ are small, they present a good accuracy result of $\Lambda_b$ decay. However, considering the $j_{\Lambda_c}^2$ type interpolating current, the uncertainties become larger. This is because the uncertainty of the input parameter of $\Lambda_b$ baryon light-cone distribution amplitudes $\epsilon_0$ has a big influence on the form factors while the other input parameters only have a small influence on the form factors and the decay properties of $\Lambda_b^0 \to \Lambda_c^+ \ell^- \overline{\nu}_{\ell}$. Taking the large uncertainties caused by the interpolating current type $j_{\Lambda_c}^2$ in the branching fractions and the comparision with form factors calculated in other models introduced in the introduction into consideration, choosing $j_{\Lambda_c}^1$ type interpolating current may be preferable.
     
     Testing the ratio of branching fractions $\mathcal{R}(\Lambda_c^+)$ of tau lepton and light lepton final states, provides an excellent way to test the Standard Model. The difference between the experiment and the Standard Model may imply that new physics beyond the Standard Model can be found. However, there are still some uncertainties about the semileptonic decay processes $\Lambda_b \to \Lambda_c \ell \overline{\nu}_\ell$. Fortunately, the recent LHCb result is consistent with the Standard Model under the uncertainty of the measurement of branching fraction $\mathcal{B}r(\Lambda_b^0\to\Lambda_c^+\mu^-\overline{\nu}_\mu)$. Even so, more precise experiments and theoretical analysis should be conducted so to explore whether there is new physics beyond Standard Model.

      \section*{Acknowledgments}
         
         This work was supported in part by the National Natural Science Foundation of China under Grant No. 11675263. We thank Prof. Yu-Ming Wang for the helpful discussions.
        
      \section*{Appendix} \label{appendix}
      	\appendix
      	\setcounter{equation}{0}
      	\renewcommand\theequation{\arabic{equation}}
         
         Light-cone QCD sum rules of $\Lambda_b \to \Lambda_c \ell \nu_{\ell}$ within the $\Lambda_c$ baryon current type $j_{\Lambda_c}^2=\epsilon_{ijk}[u^{iT}(x)C\gamma_5\gamma_\nu d^j(x)]\gamma^\nu c^k(x)$.
         
       QCD representation of correlation function with current type $j_{\Lambda_c}^2$
         \begin{align}
         T_\mu(p,q)=&-\int_{0}^{1}du\int_{0}^{\infty}d\omega \omega \tilde{\psi}_2(\omega,u)\frac{f_{\Lambda_b}^{(2)}}{(p-\omega v)^2-m_c^2}[(-\sigma M_{\Lambda_b}+\frac{M_{\Lambda_c}^2-q^2}{M_{\Lambda_b}}-m_c)\gamma_\mu \notag \\& +2v_\mu\slashed{q}-2q_\mu +\gamma_\mu\slashed{q}+2m_cv_\mu-(-\sigma M_{\Lambda_b}+\frac{M_{\Lambda_c}^2-q^2}{M_{\Lambda_b}} +m_c)\gamma_\mu\gamma_5-2v_\mu\slashed{q}\gamma_5  \notag \\& -2q_\mu\gamma_5+\gamma_\mu\slashed{q}\gamma_5 +2m_cv_\mu\gamma_5]u_{\Lambda_b}(v)
         \notag \\& +\int_{0}^{1}du\int_{0}^{\infty}d\omega\int_{0}^{\omega}d\omega'\omega'[\tilde{\psi}_2(\omega',u)-\tilde{\psi}_4(\omega',u)]\frac{f_{\Lambda_b}^{(2)}}{(p-\omega v)^2-m_c^2}(\gamma_\mu-\gamma_\mu\gamma_5)u_{\Lambda_b}(v) \notag \\& -\int_{0}^{1}du\int_{0}^{\infty}d\omega\int_{0}^{\omega}d\omega'\omega'[\tilde{\psi}_2(\omega',u)-\tilde{\psi}_4(\omega',u)]\frac{f_{\Lambda_b}^{(2)}}{[(p-\omega v)^2-m_c^2]^2}[(\overline{\sigma}M_{\Lambda_c}^2-\sigma\overline{\sigma}M_{\Lambda_b}^2 \notag \\& +\sigma q^2-m_c\overline{\sigma}M_{\Lambda_b})\gamma_\mu+2m_c\overline{\sigma}M_{\Lambda_b}v_\mu-2m_cq_\mu+m_c\gamma_\mu\slashed{q}-(\overline{\sigma}M_{\Lambda_c}^2-\sigma\overline{\sigma}M_{\Lambda_b}^2+\sigma q^2 \notag \\& +m_c\overline{\sigma}M_{\Lambda_b})\gamma_\mu\gamma_5-2m_c\overline{\sigma}M_{\Lambda_b}v_\mu\gamma_5+2m_c q_\mu\gamma_5-m_c\gamma_\mu\slashed{q}\gamma_5]u_{\Lambda_b}(v).
         \end{align}
         As the same procrdure in the current $j_{\Lambda_c}^1(x)$, and extracting the contribution of negative parity baryon $\Lambda_c^*$, we have the relations of form factors $f_1(q^2)=g_1(q^2), f_2(q^2)=f_3(q^2)=g_2(q^2)=g_3(q^2)$.
         
         Form factors $f_1(q^2)$ and $f_2(q^2)$ with the current type $j_{\Lambda_c}^2$ are

         \begin{align}
         f_1(q^2)=&-\frac{M_{\Lambda_b}}{M_{\Lambda_c}+M_{\Lambda_c^*}} \int_0^1 du \int_0^{\sigma_0}\frac{d\sigma}{\overline{\sigma}} \{\sigma[M_{\Lambda_c}(M_{\Lambda_c^*}  -M_{\Lambda_c} -m_c)-\overline{\sigma}M_{\Lambda_b}^2  \notag \\& +q^2]\psi_2(\omega,u) +2[\overline{\psi}_2(\omega,u) -\overline{\psi}_4(\omega,u)]\}e^{(M_{\Lambda_c}^2-s)/M_B^2} \notag \\& +\frac{M_{\Lambda_b}}{(M_{\Lambda_c}+M_{\Lambda_c^*})M_B^2}\int_0^1 du \int_0^{\sigma_0}\frac{d\sigma}{\overline{\sigma}^2}[(\sigma M_{\Lambda_c}-M_{\Lambda_c^*})m_c  +\sigma(\overline{\sigma}M_{\Lambda_b}^2-q^2) \notag \\& -\overline{\sigma}M_{\Lambda_c}^2][\overline{\psi}_2(\omega,u) -\overline{\psi}_4(\omega,u)] e^{(M_{\Lambda_c}^2-s)/M_B^2} \notag \\& +\frac{M_{\Lambda_b}}{M_{\Lambda_c}+M_{\Lambda_c^*}}\int_0^1 du \frac{\eta(\sigma_0,q^2)}{\overline{\sigma}_0}[(\sigma_0 M_{\Lambda_c}-M_{\Lambda_c^*})m_c  +\sigma_0(\overline{\sigma}_0 M_{\Lambda_b}^2-q^2) \notag \\& -\overline{\sigma}_0 M_{\Lambda_c}^2][\overline{\psi}_2(\omega_0,u) -\overline{\psi}_4(\omega_0,u)] e^{(M_{\Lambda_c}^2-s_0)/M_B^2},
         \end{align}
         
         \begin{align}
         f_2(q^2)=&-\frac{M_{\Lambda_b}^2}{M_{\Lambda_c}+M_{\Lambda_c^*}}\int_0^1 du \int_0^{\sigma_0}\frac{d\sigma}{\overline{\sigma}}\sigma(M_{\Lambda_c^*}-m_c)\tilde{\psi}_2(\omega,u) e^{(M_{\Lambda_c}^2-s)/M_B^2} \notag \\& +\frac{M_{\Lambda_b}^2}{(M_{\Lambda_c}+M_{\Lambda_c^*})M_B^2}\int_0^1 du \int_0^{\sigma_0} \frac{d\sigma}{\overline{\sigma}^2}\sigma M_{\Lambda_b}^2m_c[\overline{\psi}_2(\omega,u)-\overline{\psi}_4(\omega,u)]e^{(M_{\Lambda_c}^2-s)/M_B^2} \notag \\& +\frac{M_{\Lambda_b}^2}{M_{\Lambda_c}+M_{\Lambda_c^*}}\int_0^1 du \frac{\eta(\sigma_0,q^2)}{\overline{\sigma}_0^2}\sigma_0 M_{\Lambda_b}^2m_c[\overline{\psi}_2(\omega_0,u)-\overline{\psi}_4(\omega_0,u)]e^{(M_{\Lambda_c}^2-s_0)/M_B^2}
         \end{align}
         The Borel transformation
         \begin{gather}
         \int_{0}^{\infty}d\sigma\frac{\rho_i(\sigma)}{[(p-\omega v)^2-m_c^2]^2}\to \int_{0}^{\sigma_0}d\sigma\frac{1}{\overline{\sigma}^2}\frac{\rho_i(\sigma)}{M_B^2}e^{-s/M_B^2}+\frac{1}{\overline{\sigma}_0^2}\eta(\sigma_0)\rho_i(\sigma_0)e^{-s_0/M_B^2}
         \end{gather}
         is used in the above equations, where 
         \begin{align}
          \eta(\sigma)=\frac{d\sigma}{ds}=\frac{\overline{\sigma}^2}{\overline{\sigma}^2M_{\Lambda_b}+m_c^2-q^2}
         \end{align}
         and $\omega_0=\sigma_0 M_{\Lambda_b}$.
         \bibliographystyle{unsrt}
         \bibliography{lambdabRe4}

    \end{document}